\documentclass[11pt]{article}
\usepackage{amssymb,amsmath}
\usepackage{float}
\usepackage{color,fullpage}
\usepackage{amsthm}
\usepackage{graphicx}   
\usepackage{algorithm}
\usepackage{algorithmic} 
\usepackage{graphics} 
\usepackage{caption}
\usepackage{subcaption}
\usepackage{multirow}
\usepackage{url}

\newtheorem{theorem}{Theorem}

\newtheorem{lemma}{Lemma}

\numberwithin{equation}{section}
\numberwithin{figure}{section}
\numberwithin{definition}{section}

\begin{document}
\bibliographystyle{unsrt}
\title{Phase Retrieval via Sparse Wirtinger Flow}

\author{Ziyang Yuan\thanks{
College of Science, National University of Defense Technology,
Changsha, Hunan, 410073, P.R.China. Corresponding author. Email: \texttt{yuanziyang11@nudt.edu.cn}}
\and Qi Wang{\thanks{
		College of Mathematics and Statistics, Shandong Normal University,
		Jinan, Shandong, 250358, P.R.China. Email: \texttt{wangqi15@stu.sdnu.edu.cn}}
}
\and Hongxia Wang{\thanks{
		College of Science, National University of Defense Technology,
		Changsha, Hunan, 410073, P.R.China. Email: \texttt{wanghongxia@nudt.edu.cn}}
}
}



\date{}

\maketitle

\begin{abstract}
	Phase retrieval(PR) problem is a kind of ill-condition inverse problem which can be found in various of applications. Utilizing the sparse priority, an algorithm called SWF(Sparse Wirtinger Flow) is proposed in this paper to deal with sparse PR problem based on the Wirtinger flow method.  SWF firstly recovers the support of the signal and then updates the evaluation by hard thresholding method with an elaborate initialization. Theoretical analyses show that SWF has a geometric convergence for any $k$ sparse $n$ length signal with the sampling complexity $\mathcal{O}(k^2\mathrm{log}n)$. To get $\varepsilon$ accuracy, the computational complexity of SWF is   $\mathcal{O}(k^3n\mathrm{log}n\mathrm{log}\frac{1}{\varepsilon})$.
	Numerical tests also demonstrate that SWF performs better than state-of-the-art methods especially when we have no priori knowledge about sparsity $k$. Moreover, SWF is also robust to the noise\\
	$\mathbf{keywords}$:~~sparse phase retrieval,~~Wirtinger flow,~~gradient descent,~~hard thresholding
	
\end{abstract}

\section{Introduction}
\label{intro}
\subsection{Phase retrieval problem}
\indent 
 Recovering a signal from its intensity only measurements is called phase retrieval(PR) problem arised in a wild range of applications such as Fourier Ptychography Microscopy, diffraction imaging, X-ray crystallography and so on\cite{Bunk2007Diffractive}\cite{Miao1999Extending}\cite{Zheng2013Wide}. PR problem can be an instance of solving a system of quadratic equations: 
\begin{eqnarray}
\centering
y_i=|\mathbf{a}_i^*\mathbf{x}|^2+\mathbf{\varepsilon}_i,~~i=1,...,m,
\end{eqnarray}
where $\mathbf{x}\in\mathbb{C}^n$ is the signal of interest, $\{\mathbf{a}_i\}_{1\leq i\leq m}\in\mathbb{C}^n$ are the measurement vectors, $\mathbf{y}=[y_1,y_2,...,y_m]^{T}\in \mathbb{R}^{m}$ is the observed measurement, vector $\mathbf{\varepsilon}=[\varepsilon_1,\varepsilon_2,...,\varepsilon_n]^T$ is the noise.\\ 
\indent
(1.1) is a non-convex and NP-hard problem. Traditional methods usually fail to find the solutions. Besides, let $\tilde{\mathbf{x}}$ be the solution of (1.1), $\tilde{\mathbf{x}}e^{j\theta}$ also satisfies (1.1) for any $\theta\in[0,2\pi)$ where $j=\sqrt{-1}$. So the uniqueness of the solution of (1.1) is often defined up to a global phase factor.
\subsection{Prior art}
For classical PR problem, $\{\mathbf{a}_i\}_{1\leq i\leq m}$ are the Fourier measurement vectors. There were series of methods to solve (1.1). In 1970, error reduction methods such as Gerchberg-Saxton and Hybrid input and output method\cite{Fienup1982Phase}\cite{Gerchberg1971A} were proposed to tackle with PR by constantly projecting the evaluations between the transform domain and the spatial domain with some special constraints. These methods often get stuck into the local minimums. In addition, fundamental mathematical questions concerning their convergence remain unsolved. In fact, without any additional assumption about $\mathbf{x}$, it is hard to recover $\mathbf{x}$ from $\{y_i\}_{1\leq i\leq m}$. For Fourier measurement vectors, the trivial ambiguities of (1.1) include global phase shift, conjugate inversion and spatial shift. In fact, it has been proven that 1D Fourier phase retrieval problem has no unique solution even excluding those trivialities above. To relief those ill-condition characters, one way is to substitute Fourier measurements with other measurements owning the high redundancy property like Gaussian measurements\cite{Cand2013PhaseLift}\cite{Waldspurger2012Phase}\cite{candes2015phase}, coded diffraction pattern\cite{Cand2013Phase}, wavelet frame\cite{Mallat2015Phase} and so on.\\
\indent
Sparsity priori has been considered in many problems with significant application meanings. Greedy sparse phase retrieval(Gespar) is a kind of heuristics to deal with phase retrieval problem\cite{Shechtman2013GESPAR} based on but not restricted to Fourier measurements. It utilizes the damped Gaussian-Newton algorithm to search for the local minimum and then updates the support of signal by 2-opt method. In\cite{Cand2013PhaseLift}\cite{Waldspurger2012Phase}\cite{Candes2011Phase}, SDP(Semi-Definite Programming) based algorithms were came up to deal with phase retrieval problem by lifting (1.1) into a higher dimension. This convex alternative can deal with phase retrieval problem but need high computational costs. Plugging the $l_1$ constraint into this convex objective function, Compressive Sensing Phase Retrieval via Lifting(CPRL) was proposed to tackle with the sparse phase retrival. On the other hand, to decrease the computational cost,  Alternating minimization(ALTMIN)\cite{Netrapalli2013Phase} method alternativly updates signal and phase to search for the signal of interest. It can deal with sparse signal by adding hard-thresholding method in each iteration.\\
\indent
Another wildspread nonconvex method for phase retrival is the WF(Wirtinger flow)\cite{candes2015phase} which directly utilizes the gradient descent method to search for the global minimum. Based on WF, there are several variants under different conditions. For Poisson likelihood function, truncated Wirtinger flow(TWF) was proposed\cite{chen2015solving}. For amplitude based models, reshaped WF and TAF(Truncated Amplituded Flow) were considered in \cite{zhang2016reshaped}\cite{wang2016solving}. Those methods often need $m/n>3$ for the exactly recovery. But, on the theoretical side, for a $k$ sparse signal $\mathbf{x}\in\mathbb{R}^n$, $4k-1$ measurements are sufficient to guarantee uniqueness. The gap of the sampling complextiy is large. Thus, it is necessary to come up with an algorithm for sparse wirtinger flow phase retrival problem. In\cite{wang2016solving1}\cite{Cai2016Optimal}, thresholding WF and Sparse Phase Retrieval via Truncated Amplitude(SPARTA) were camp up utilizing the priority of sparsity. SPARTA can have a high recovery rate than thresholding WF with a faster converge rate. But SPARTA is sensitive to the priority $k$.\\
\indent
Based on the thresholding WF and SPARTA, A sparse phase retrieval problem called SWF is proposed in this paper. We adopt the Gaussian maximum likelihood function as the objective which is a forth order smooth function with a benign geometrical property. Then we use a two-stage algorithm to find the global optimum. In the first stage, the support of the signal is estimated by a well justified rule in \cite{Wang2016Sparse}, then we apply the truncated spectral method to evaluate the initialization which is restricted to the support evaluated above. In the second stage, the initialization is constantly refined by the hard thresholding based gradient descent method. The sample complexity and computational complexity of SWF can be seen in table 1. Theoretical results show that SWF can recover any $k$ sparse $n$ dimension signal $\mathbf{x}$ through $\mathcal{O}(k^2\mathrm{log}n)$ measurements with the minimum nonzero entries's modulus on the order of $\frac{1}{k}||\mathbf{x}||$. Besides, SWF have a geometric convergence rate and need $\mathcal{O}(k\mathrm{log}\frac{1}{\varepsilon})$ iterations to get $\varepsilon-$ accuracy. Though the computational complexity of SWF is $k$ times larger than SPARTA, numerical tests show that the SWF can have a better performance than state of the art methods. Specifically, SWF can have a 100\% recovery rate when $m=0.7n$. Especially when the priori sparsity $k$ is unkown, SWF is also significantly superior to other algorithms. Moreover, SWF can be insensitive to the support misspecification and noise in some degree. For readers to be convenient to replicate the simulating tests, the codes of SWF are available at:\url{https://github.com/Ziyang1992/Sparse-Wirtinger-flow.git}\\   
\begin{table}
		\centering
	\begin{tabular}{|c|c|c|}
		\hline
		~~&Sample complexity&Computational complexity\\
		\hline
		CPRL\cite{Ohlsson2012CPRL}& $\mathcal{O}(k^2\mathrm{log}n)$&$\mathcal{O}(n^3/\varepsilon^2)$\\	
		\hline	
		ALTMIN\cite{Netrapalli2013Phase} &$\mathcal{O}(k\mathrm{log}n(k+\mathrm{log}^3k+\mathrm{log}\frac{1}{\varepsilon}\mathrm{log}\mathrm{log}\frac{1}{\varepsilon})$& $\mathcal{O}(k^2\mathrm{log}n(kn+\mathrm{log}^2\frac{1}{\varepsilon}\mathrm{log}\mathrm{log}\frac{1}{\varepsilon})$\\
		\hline
  		Thresholding WF\cite{Cai2016Optimal}&$\mathcal{O}(k^2\mathrm{log}n)$&$\mathcal{O}(k^3n\mathrm{log}n\mathrm{log}\frac{1}{\varepsilon})$\\
		\hline
	SPARTA\cite{Wang2016Sparse}&$\mathcal{O}(k^2\mathrm{log}n)$&$\mathcal{O}(k^2n\mathrm{log}n\mathrm{log}\frac{1}{\varepsilon})$\\
	\hline
	 SWF&$\mathcal{O}(k^2\mathrm{log}n)$&$\mathcal{O}(k^3n\mathrm{log}n\mathrm{log}\frac{1}{\varepsilon})$\\
	\hline
	\end{tabular}
	\caption{Comparison of State of art approach}
\end{table}  
\subsection{Contribution of this paper}
\indent
Our contributions are in two folds. Firstly, we propose an algorithm called SWF to solve sparse phase retrieval problem. Though the amplitude based algorithm can be superior to the intensity based algorithm for the general signal when the ratio $m/n$ is small\cite{zhang2016reshaped}\cite{wang2016solving}. In this paper, we find that when it comes to the sparse signal, the intensity based model ——SWF did have a better performance than the amplitude based one.
The second is our theoretical contribution. We prove that SWF has a linear convergence with measurements $m$ largely exceeding the sparsity $k$.\\  
\indent
The remainders of this paper are organized as follows. In section 2, we introduce the proposed SWF and establish its theoretical frames. In section 3, numerical tests compare SWF with state-of-the-art approaches. Section 4 is the conclusion. Technical details can be found in Appendix.\\
\indent 
In this article, the bold uppercase and lowercase letters represent matrices and vectors. $(\cdot)^{T}$ denotes the transpose. $|\cdot|$ denotes the absolute value of a real number or the cardinality of a set. $||\cdot||$ is the Euclidean norm of a vector. $||\cdot||_0$ is the zero norm.
\section{Sparse Wirtinger Flow}
\subsection{Algorithm of SWF}
Sparse phase retrieval aims to find an evaluation $\mathbf{z}$ approximating to $k$ sparse signal $\mathbf{x}$ from (1.1),
\begin{eqnarray}
\nonumber&\mathrm{Find}~~\mathbf{z}&\\
\nonumber&s.t.~~ y_i=|\mathbf{a}_i^T\mathbf{z}|^2+\mathbf{\varepsilon}_i,~~i=1,...,m,&\\
&||\mathbf{z}||_0=k.&
\end{eqnarray} 
In our paper, the sparstiy $k$ is assumed to be known as a priori for the theorectical simplicity. We also make simulation tests to show the performance of SWF when the priori sparsity $k$ is unknown. \\
We assume that $\mathbf{a}_i\sim\mathcal{N}(\mathbf{0},\mathbf{I})$ and $\varepsilon_i\sim\mathcal{N}(0,\sigma)$, then the probability density function of $\varepsilon_i$ is:
\begin{eqnarray}
\mathbb{P}(\varepsilon_i)=\frac{1}{\sqrt{2\pi}\sigma}\mathrm{exp}(-\frac{(y_i-|\mathbf{a}_i^T\mathbf{z}|^2)^2}{2\sigma^2}).
\end{eqnarray}
According to (2.2), neglecting the effects of constants and assume the signal is real, we estimate the maximum likelihood function as
\begin{eqnarray}
\nonumber&\mathop {\mathrm{minimize}}\limits_{\mathbf{z}\in\mathbb{C}^n}~~\mathit{f}(\mathbf{z})=\frac{1}{2m}\sum_{i=1}^{m}\big((\mathbf{a}_i^T\mathbf{z})^2-y_i\big)^2,&\\
&s.t.~~||\mathbf{z}||_0=k.&
\end{eqnarray}
(2.3) is a non-convex optimization problem which has many local minimums. As a result, it seems impossible to solve (2.3) with convex methods. Owning to the statistical property of Gaussian random vectors, (2.3) can have a benign geometrical structure. Without the sparisty constraint, a butch of algorithms were came up to search for the global optimums\cite{candes2015phase}\cite{gao2016gauss}\cite{sun2016geometric}\cite{li2016gradient}. Those method performs well when $m/n$ is large enough. We can utilize the sparse condition to decrease the sample complexity. Thus for sparse signal, we come up with a more efficient algorithm called SWF. Firstly, we recover the support of $\mathbf{x}$, then we apply the truncated spectral method to make a good initialization under the recovered supports. At last, we utilize a hard thresholding based gradient descent algorithm to search for the global optimum. Next, we will give the details of SWF without noise from these three parts.\\
\indent
Before explanation, we introduce several notations. The distance between the evaluation $\mathbf{z}$ and real solution $\mathbf{x}$ is defined as:
\begin{eqnarray*}
\mathrm{dist}(\mathbf{z},\mathbf{x})=\mathop {\mathrm{min}}\limits_{\phi\in[0,2\pi)}||\mathbf{z}-\mathbf{x}e^{j\phi}||,
\end{eqnarray*}
where $j=\sqrt{-1}$.\\
\indent
Then, for any vector $\mathbf{z}$ and any support $S$, $\mathbf{z}_S$ means vector $\mathbf{z}$ deletes all the elements outside of support $S$. 
\subsubsection{Support recovery}
\indent
To recover the support of $\mathbf{x}$, we use the same method in \cite{Wang2016Sparse}. Assuming $\mathbf{x}\in\mathbb{R}^n$ is a $k$ sparse signal with support $S^*$, $|S^*|=k$. We define $D_{i,j}=(\mathbf{a}_i^{T}\mathbf{x})^2a_{i,j}$ where $a_{i,j}$ is the $j$th element of $\mathbf{a}_i$. Note that $\mathbf{a}_i\sim\mathcal{N}(\mathbf{0},\mathbf{I})$, by calculating the moment of Gaussian variables, we have $\mathbb{E}(|a_{i,j}|^4)=3$, $\mathbb{E}(|a_{i,j}|^2)=1$.\\
\indent
So,
\begin{eqnarray}
\mathbb{E}(D_{i,j})=\mathbb{E}\big((\sum_{k=1}^{n}a_{i,k}x_k)^2a_{i,j}\big)=||\mathbf{x}||^2+2x_j^2.
\end{eqnarray}
Denote $E_j=\frac{1}{m}\sum_{i=1}^{m}D_{i,j}$.
From (2.4), the differences between $E_j$ are determined by $x_j$ if $m$ is large enough. If $x_j^2$ is larger, accordingly the $E_j$ is larger too. Then we can sort out the $k$ largest $E_j$ and record their indexes as the estimated support $S_0$. Lemma 1 guarantees the accuracy of this support recovery method.\\
\begin{lemma}\cite{Wang2016Sparse}
	For any $k$ sparse signal $x\in\mathbb{R}^n$ with support $S^*$ and minimum nonzero entries $x_{\mathrm{min}}:=\mathrm{min}_{j\in S^*}|x_j|$ on the order of $(1/\sqrt{k})||\mathbf{x}||_2$. If $\mathbf{a}_i\overset{i.i.d}\sim\mathcal{N}(\mathbf{0},\mathbf{I}_n)$, $i=1,...,m$. $S_0$ is equal to $S^*$ with a probability at least $1-6/m$ provided $m\geq C_0k^2\mathrm{log}(mn)$ for some constant $C_0$.
\end{lemma}   
\indent
Lemma 1 shows that when $m$ is sufficiently large, $S_0$ approximates $S^*$ quite well. But from numerous tests, we find that SWF can still recover $\mathbf{x}$ even when $S_0$ is quite different from $S^*$.\\
\subsubsection{Initialization evaluation}
We have estimated the support $S_0$ of $\mathbf{x}$, now we constrain $\mathbf{a}_i$ on $S_0$, i.e., deleting those elements which aren't in $S_0$. Under the guarantee of lemma 1, we will use the truncated spectral method to make an initialization. Specifically, we construct a matrix $\mathbf{Y}$ as (2.5). When $m$ is sufficiently large, the eigenvector of $\mathbf{Y}$ can be taken as an approximation of  $\mathbf{x}$,
\begin{eqnarray}
	\mathbf{Y}=\frac{1}{m}\sum_{i=1}^{m}y_i\mathbf{a}_{i,S_0}\mathbf{a}_{i,S_0}^T\mathbf{1}_{\{|y_i|\leq\alpha^2_y\phi^2\}},
\end{eqnarray}
where $\alpha_y$ is the truncation threshold. $\phi^2=\frac{1}{m}\sum_{i=1}^{m}y_i$.\\
\indent
 Lemma 2 demonstrates the accuracy of the estimation made by the truncated spectral method.  
\begin{lemma}
	Under the conditions of lemma 1, for any $\delta>0$ and $\mathbf{x}\in\mathbb{R}^n$, the solution $\mathbf{z}^{0}_{S_0}\in\mathbb{R}^k$ returned by the truncated spectral method obeys:
	\begin{eqnarray}
	\mathrm{dist}(\mathbf{z}^{0}_{S_0},\mathbf{x}_{S_0})\leq\delta||\mathbf{x}_{S_0}||.
	\end{eqnarray}
	with probability not less than $1-\mathrm{exp}(-C_1m)$, providing that $m>c_0k$ for some constant $C_1$ and $c_0>0$ which is determined by $\delta$.	
\end{lemma}

\noindent
\emph{Proof}:\\
\indent
Based on the condition of lemma 1, we have $S_0=S^*$ with a probability at least $1-\frac{6}{m}$ provided $m\geq C_0k^2\mathrm{log}(mn)$. Then (2.5) can be rewrited as
\begin{eqnarray}
\frac{1}{m}\sum_{i=1}^{m}(\mathbf{a}_i^{T}\mathbf{x})^2\mathbf{a}_{i,S_0}\mathbf{a}_{i,S_0}^{T}=\frac{1}{m}\sum_{i=1}^{m}(\mathbf{a}_{i,S^*}^{T}\mathbf{x}_{S^*})^2\mathbf{a}_{i,S_0}\mathbf{a}_{i,S_0}^{T}=\frac{1}{m}\sum_{i=1}^{m}(\mathbf{a}_{i,S_0}^{T}\mathbf{x}_{S_0})^2\mathbf{a}_{i,S_0}\mathbf{a}_{i,S_0}^{T}.
\end{eqnarray}
Combining with proposition 3 in \cite{chen2015solving}, we can conclude lemma 2 is true.\\
\indent
Utilizing power method in (2.5) to get the estimation $\mathbf{z}^{0}_{S_0}\in\mathbb{R}^k$, we will scale  $||\mathbf{z}^0_{S_0}||=\phi$ and construct $\mathbf{z}_0\in\mathbb{R}^n$ where elements in $S_0$ are equal to $\mathbf{z}^{0}_{S_0}$, others are all zero. In the test, we generally set $\alpha_y=3$ and run the power method with 100 iterations.
\subsubsection{Hard thresholding based gradient descent}
We utilize the hard-thresholding based gradient descent algorithm to search for the global optimum in each iteration with $\mathbf{z}_0$ as an initialization.\\
The gradient of $f(\mathbf{z})$ is calculated by the Wirtinger derivative.
\begin{eqnarray}
\nabla f(\mathbf{z})=\frac{2}{m}\sum_{i=1}^{m}\big((\mathbf{a}_i^T\mathbf{z})^2-y_i^2\big)\mathbf{a}_i\mathbf{a}_i^T\mathbf{z}.
\end{eqnarray} 
In the $t$th iteration of gradient descent, we have:
\begin{eqnarray}
\tilde{\mathbf{z}}^{t}=\mathbf{z}^{t-1}-\frac{\mu_{t}}{\phi^2}\nabla f(\mathbf{z}^{t-1}),
\end{eqnarray}
where $\mu_t$ is the step size.
Here, we add a thresholding operator $\mathcal{T}_k$ to $\tilde{\mathbf{z}}^{t}$.
\begin{eqnarray}
\mathcal{T}_k(\tilde{\mathbf{z}}^{t})=\mathbf{z}^t,
\end{eqnarray}
where $\mathbf{z}^t$ keeps the $k$-largest absolute value of $\tilde{\mathbf{z}}^{t}$ and sets other elements to zero. $\mathcal{T}_k$ projects $\tilde{\mathbf{z}}^{t}$ into the subspace $\mathbb{V}_k=\big\{\mathbf{v}\in\mathbb{R}^n\big|||\mathbf{v}||_0\leq k\big\}$. This procedure can decrease the freedom dimension and constrain the searching domain. Numerous tests also show hard thresholding procedure is effective for sparse PR. Theorem 1 guarantees the convergence of this hard thresholding based gradient descent method. 
\begin{theorem}
	Based on lemma 1 and lemma 2, with a proper stepsize $\mu_t$, the $t+1$th estimation of $\mathrm{SWF}$ $\mathbf{z}^{t+1}$ satisfies:
	\begin{eqnarray}
	||\mathbf{z}^{t+1}-\mathbf{x}||\leq\delta_1 (1-\nu)^{t+1}||\mathbf{x}||,
	\end{eqnarray}
	with probability exceeding $1-c_1m^{-1}-c_2\mathrm{exp}(-c_3k/\mathrm{log}m)$ provided $m\geq C_0 k\mathrm{log}k$, where $0<\nu<1$, $C_0$, $c_1$, $c_2$, $\delta_1$ are all constants. 
\end{theorem}
The proofs of theorem 1 are in appendix. The stepsize $\mu_t$ can be $0.1$ for all $t$, then $\nu=0.19$. In our simulation tests, in order to make SWF have a good performance when $m$ isn't larege enough, we select a varying stepsize. The stepsize $\mu_t=\mathrm{min}\Big(\big(1-\mathrm{exp}(\frac{-t}{330})\big)/2,0.1\Big)$ is utilized in the SWF. In the first few iterations, the stepsize is small to prevent iteration from stagnating into the local minimum easily, then we gradually increase the value of stepsize. Combining lemma 1, lemma 2 and theorem 1, we can get the exact recovery guarantee for SWF. The details of SWF can be seen clearly in algorithm 1.\\
\begin{algorithm}
	\caption{\textbf{Sparse Wirtinger Flow}($\mathbf{SWF}$)} 
	\label{alg:Framwork} %
	\renewcommand{\algorithmicrequire}{\textbf{Input:}}
	\renewcommand\algorithmicensure {\textbf{Output:} }
	\begin{algorithmic}   
		\REQUIRE$\{\{y_i\}_{1\leq i\leq m},\{\mathbf{a}_i\}_{1\leq i\leq m},T,k,\mu_t,\alpha_y\}$ 
		\STATE 
		$\{\mathbf{a}_i\}_{i=1}^m$: Gaussian vectors\\
		$y_i=|\langle\mathbf{a}_i,\mathbf{x}\rangle|^2$: measurements\\
		$\varepsilon$: the accuracy required\\
		$k$: the sparsity of $\mathbf{x}$\\
		$\mu_t$: the step size\\
		$\alpha_y$: truncation thresholds\\
		$T$: the maximum iteration times\\
		\ENSURE$\hat{\mathbf{x}}$
		\STATE $\hat{\mathbf{x}}$: the estimated signal\\
		\vskip 4mm
		\hrule
		\vskip 2mm
		$\mathbf{Support~recovery}$
		\STATE
		set $S_0$ to be the set of $k$ largest indices of $\{\frac{1}{m}\sum_{i=1}^{m}y_ia_{i,j}^2\}_{1\leq j\leq n}$\\
		\textbf{Initialization evaluation}
		\STATE
		$\phi^2=\frac{1}{m}\sum_{i=1}^{m}y_i$\\
		Let $\mathbf{z}^0_{S_0}(||\mathbf{z}^{0}_{S_0}||=\phi)$ to be the eigenvector corresponding to the largest eigenvalue of
		\begin{eqnarray*}
			\mathbf{Y}=\frac{1}{m}\sum_{i=1}^{m}y_i\mathbf{a}_{i,S_0}\mathbf{a}_{i,S_0}^*\mathbf{1}_{\{|y_i|\leq\alpha^2_y\phi^2\}}
		\end{eqnarray*}
	   set $\mathbf{z}^0$ to be the vector where elements in $S_0$ are equal to $\mathbf{z}^{0}_{S_0}$, others are all zero.\\
	  \textbf{Hard thresholding based gradient descent}\\
	  $t=1$, $\mathbf{z}^1=\mathcal{T}_k(\mathbf{z}^{0}-\frac{\mu_0}{\phi^2}\nabla\mathit{f}(\mathbf{z}^0))$
		\WHILE{$||\mathbf{z}^{t}-\mathbf{z}^{t-1}||\geq\varepsilon$ and $t\leq T$}
		\STATE
	$\mathbf{z}^{t+1}=\mathcal{T}_k(\mathbf{z}^{t}-\frac{\mu_t}{\phi^2}\nabla\mathit{f}(\mathbf{z}^t))$\\
	\STATE $t=t+1$
		\ENDWHILE\\
		$\mathbf{\hat{x}}=\mathbf{z}^t$
	\end{algorithmic}  
\end{algorithm}
\indent
In the next section, we will make several simulations to demonstrate the effectiveness of SWF with the comparison of state of the art methods.\\
\section{Numerical tests}
Numerical results are given in this section which show the performance of SWF together with SPARTA\cite{Wang2016Sparse}, ALTMIN\cite{Netrapalli2013Phase}, TAF\cite{wang2016solving} and Thresholding WF\cite{Cai2016Optimal}. All the tests are carried out on the Lenovo desktop with a 3.60 GHz Intel Corel i7 processor and 4GB DDR3 memory. Here,  we are in favor of normalized mean square error ($\mathbf{NMSE}$) which can be calculated as:
\begin{eqnarray*}
	\centering
	\mathbf{NMSE}=\frac{||\hat{\mathbf{x}}-\mathbf{x}||}{||\mathbf{x}||},
\end{eqnarray*}
where $\hat{\mathbf{x}}$ is the numerical estimation of $\mathbf{x}$.\\
\indent
In all simulating tests, $\mathbf{x}\in\mathbb{R}^{1000}$ is a real Gaussian random vector satisfying $\mathcal{N}(\mathbf{0},\mathbf{I})$. $\{\mathbf{a}_i\}_{1\leq i\leq m}$ are drawn from $\mathcal{N}(\mathbf{0},\mathbf{I})$. The stepsize $\mu_t=\mathrm{min}\Big(\big(1-\mathrm{exp}(\frac{-t}{330})\big)/2,0.1\Big)$ is utilized in the SWF. For all tests, if the $\mathbf{NMSE}$ is below $10^{-5}$, we will regard it as a success. The successful times divided by testing times is the recovery rate. \\  
\noindent
\textbf{Test~1} \\
\indent
In the first test, we assume the sparsity $k$ is known. The signal $\mathbf{x}$ is fixed with sparsity $k=10$. The ratio between $m$ and $n$ ranges from $0.1$ to $3$. At each ratio, we run 100 times tests. The recovery rate of different methods are shown in Figure 3.1.\\
\begin{figure}	
	\centering
	\includegraphics[width=3in]{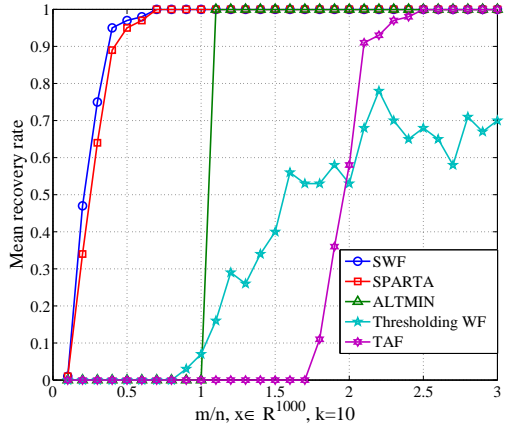}
	\caption{The comparion of different algorithms for fixed $k=10$ with different $m/n$}\label{fig:1a}		
\end{figure}
\indent
From Figure 3.1, we can see that SWF is a little superior to SPARTA. SWF and SPARTA can have a 100\% recovery rate when $m\geq0.7n$. But ALTMIN has 100\% recovery rate only when $m\geq1.1n$. TAF is one of the best algorithm for general phase retrieval which isn't designed for sparse signal. TAF~can have a 100\% recovery rate when $m\geq2.5n$. Thresholding WF can recover signal when $m/n<1$, but it can't get a 100\% recovery rate for all these ratios.\\
\noindent
\textbf{Test~2} \\
\indent
Assume the sparsity $k$ is unkown. All test settings are the same with test 1. But the $k$ we known as a priori is taken as $\sqrt{n}\approx32$ according to the sample complexity in tabel 1. The results are shown in Figure 3.2.\\
\indent
From Figure 3.2 we can find that SWF can be superior to SPARTA when the priori sparsity $k$ isn't known correctly. SWF can have a recovery rate about 90\% at $m/n=0.5$, but SPARTA nearly can't recover $\mathbf{x}$ at the same ratio. ALTMIN can still have a 100\% recovery rate when $m=1.1n$.\\
\begin{figure}	
	\centering
	\includegraphics[width=3in]{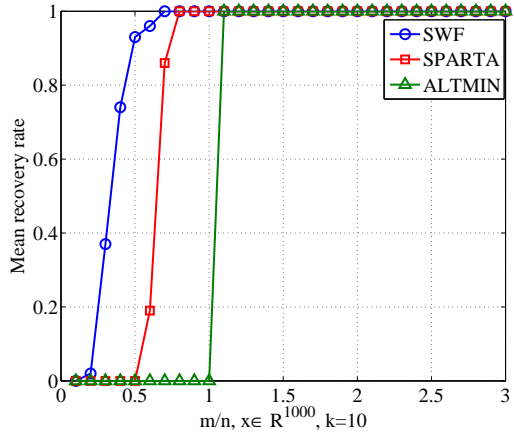}
	\caption{The comparion of different algorithm for fixed $k=10$ with different $m/n$ ratio}\label{fig:1a}		
\end{figure}
Table 2 shows the average iterations of each algorithm and corresponding time needed in test 1 and test 2. We select three different $m/n$ ratios to make comparison. We refer SWF, SPARTA and ALTMIN in test 2 as SWF0, SPARTA0 and ALTMIN0. From table 2 we can find out that the SPARTA and SPARTA0 will need fewer iterations and less time to attain the required accuracy. Because they are all based on the truncated amplitude method which is one of the most efficient algorithms for general wirtinger flow phase retrieval. But it can't have a high recovery rate comparing to SWF. Besides the time and iterations SWF need cost is considerable.\\
\begin{table}[htp]
	\centering
	\caption{Computational complexity of different methods in test 1 and test 2}
	\vspace{0.2cm}
	\begin{tabular}{cccccccc}
		\hline
		\multicolumn{1}{c}{} & \multicolumn{3}{c}{Iterations}& & \multicolumn{3}{c}{Time}\\
		\cline{2-4}\cline{6-8}
		\multicolumn{1}{c}{} & $m/n=0.5$ & $m/n=1$ & $m/n=1.5$& & $m/n=0.5$ & $m/n=1$ & $m/n=1.5$\\
		\hline
		\multirow{0}{*} SWF &79 &69 &66 & &0.360 &0.690 &1.070\\
		SPARTA&12&9&8& &0.007&0.012&0.0213\\
		ALTMIN &-&-&4& &-&- &2.267\\
		Thresholding WF &-&697&707& &-&6.953&11.784\\
		TAF &-&-&-& &-&-&-\\
		SWF0&123&90&80 & &0.575 &0.902 &1.289\\
		SPARTA0&-&24&16& &- &0.022 &0.034\\
		ALTMIN0&-&-&6& &-&-&2.675\\
		\hline
	\end{tabular}
\end{table}
\noindent
\textbf{Test~3} \\
\indent
 To find out the ability of those methods in resisting for the misspecific priori sparsity $k$, we fix $n=m$, and the sparsity of $\mathbf{x}$ is 10. The priori sparisty $k$ ranges from $5$ to $100$. At each $k$, we also run 100 tests.\\
 \indent
From Figure 3.3, we can see that the misspecification ability of SWF are better than SPARTA. Especially for SWF, the recovery rate of it can be even 95\% when the priori sparsity is $100$.\\ 
\begin{figure}	
		\centering
		\includegraphics[width=3in]{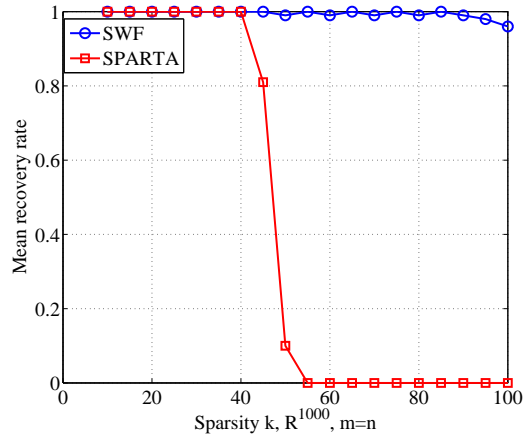}
	\caption{The comparison between SWF and SPARTA on misspecific priori sparsity.}
\end{figure}
\indent
Why the SWF and SPARTA can have diffrent results with similar procedures? We think this can be attributed to the truncated procedure in SPARTA. Because SPARTA truncates some components of gradient to make the direction of truncated gradient heading to the global minimum. But this theory is based on the condition that $m$ is sufficiently large than $k$. When this condition can't be satisfied, the truncated procedure may neglect some positive information. Here is an example where $\mathbf{x}=[1.0838,0,0]$ with sparsity $k=1$, $m=n=3$ which can be shown in Figure 3.4. The priority sparsity is $k=2$. Then we use the initialization estimated by SPARTA to recover $\mathbf{x}$ by SWF and SPARTA. The black triangle is both algorithms' initialization $\mathbf{z}_0=[0.8915, 0, -0.3987]$. Blue circles are the iterations made by SWF which converge to the point $\hat{\mathbf{x}}=[1.0838,0,-8.919\times10^{-6}]$ after 70 iterations. $\mathbf{NMSE}$ of SWF is below $10^{-5}$. But the iterations made by SPARTA which are shown by red circles are stagnated into the local minimum point $\mathbf{z}^l=[0.8688, 0, -0.4806]$.\\
\begin{figure}	
	\centering
	\includegraphics[width=3in]{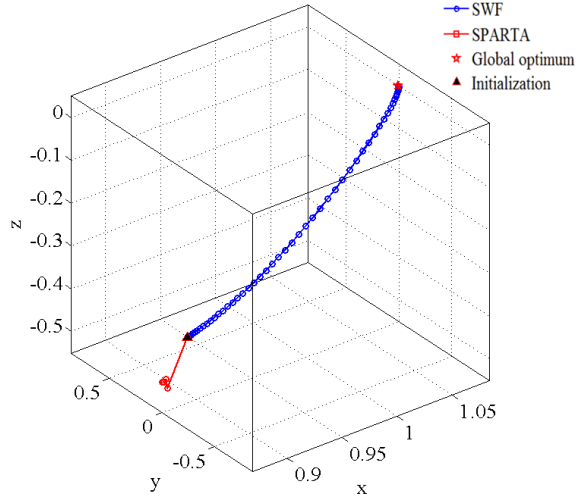}
	\caption{The itertations of SWF and SPARTA}		
\end{figure}

\noindent
\textbf{Test~4} \\
\indent
Next, we will research how the sparsity $k$ affects the recovery rate of those algorithms. Here, $m=1.5n$, the sparsity $k$ of $\mathbf{x}$ varies from 10 to 100. There is no misspecification for priori sparsity. At each sparsity $k$, we also run 100 tests. The results are shown in Figure 3.5. We can find that SWF can be superior to othe algorithms and can have a mean recovery rate about $60 $\% when sparsity $k=100$.\\ 
\begin{figure}
	\centering
	\includegraphics[width=3in]{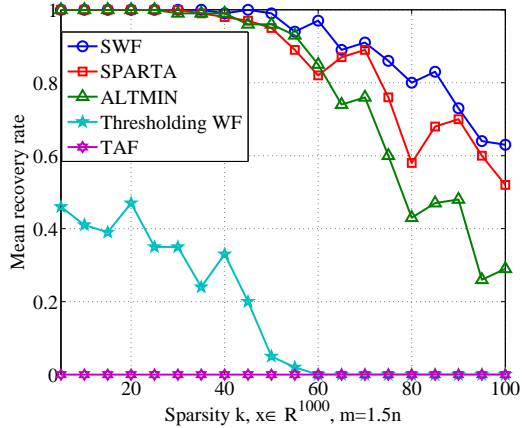}
	\caption{Comparison of different algorithm for different sparsity $k$}
\end{figure}

\noindent
\textbf{Test~5} \\
\indent
At last, we will test the robustness of different algorithms. We assume the noise model is described as (2.1). The noise is Gaussian white noise and the SNR varies from $5$dB to $10$dB. We fix $k=10$ and $m=1.5n$. At each SNR, we run 100 tests and record the average \textbf{NMSE}. The results are shown in Figure 3.6. We can see that SWF is more robust to the noise than SPARTA and ALTMIN.\\
\begin{figure}	
		\centering
		\includegraphics[width=3in]{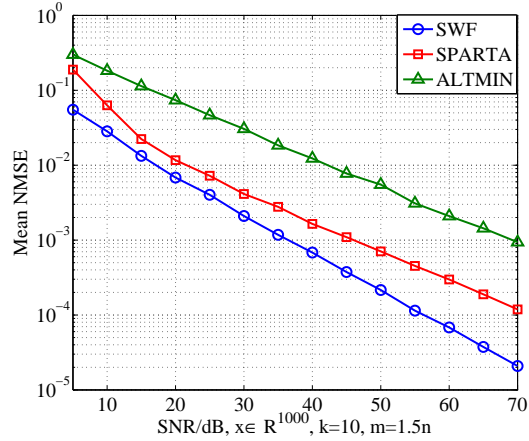}
		\caption{Comparison of different algorithm to the Nosie}		
\end{figure}
\indent
All in all, SWF has a high recovery rate than SPARTA and ALTMIN especially when the sparsity $k$ isn't known exactly. Besides it is also robust to noise. The iteration and time that SWF costs is also considerable .\\
\indent 

\section{Conclusion}
In this paper, we propose a Wirtinger flow algorithm for sparse phase retrieval problem. It can have a high recovery rate when the sampling complexity is low with support recovery and hard-thresholding based gradient descent. This algorithm aims to the 1D signal and has a good performance when the signal is real. In the future work, we will be keen to broaden it and modify it to be appropriate for high dimensional signals. 
\section{Acknowledgement}
This work was supported in part by National Natural Science foundation(China): 61571008.



\bibliography{1.bib}
\section{Appendix}
\subsection{Preliminaries}
Let $\Theta_{t+1}=S_{t+1}\cup S^*$, $S_{t+1}$ is the support of $\mathbf{z}^{t+1}$, $S^{*}$ is the support of real solution $\mathbf{x}$. The difference of two set $\Theta_{t+1}$ and $\Theta_{t}$ can be defined as $\Theta_{t+1}\setminus\Theta_{t}$. We can clearly know that $|S^*|=|S_{t+1}|=k$, $|\Theta_{t+1}|\leq2k$, $|\Theta_t\setminus\Theta_{t+1}|\leq2k$ as well as $|\Theta_{t+1}\cup\Theta_{t}|\leq3k$ for all $t$. The proof of theorem 1 will be based on \cite{Wang2016Sparse},\cite{sun2016geometric} and \cite{Cai2016Optimal}. To proof the linear convergence in theorem 1, we will get the relationship below from every iteration,
\begin{eqnarray}
||\mathbf{h}^{t+1}||\leq\nu||\mathbf{h}^t||,~~0<\nu<1,
\end{eqnarray} 
where $\mathbf{h}^t=\mathbf{z}^t-\mathbf{x}$.\\
\noindent
\textbf{Proof of Theorem 1}\\
Let $\mathbf{z}^{t+1}$ be the estimation in the $t+1$th steps of algorithm 1. With the triangle inequality, we have:
\begin{eqnarray}
||\mathbf{h}^{t+1}||=||\mathbf{z}^{t+1}-\mathbf{x}||=||\mathbf{z}^{t+1}_{\Theta_{t+1}}-\mathbf{x}_{\Theta_{t+1}}||
\leq||\mathbf{z}^{t+1}_{\Theta_{t+1}}-\tilde{\mathbf{z}}^{t+1}_{\Theta_{t+1}}||+||\mathbf{x}_{\Theta_{t+1}}-\tilde{\mathbf{z}}^{t+1}_{\Theta_{t+1}}||,
\end{eqnarray}
where $\tilde{\mathbf{z}}^{t+1}_{\Theta_{t+1}}=\mathbf{z}^{t}_{\Theta_{t+1}}-\frac{\mu_t}{\phi^2}\nabla f(\mathbf{z}^{t})_{\Theta_{t+1}}$.\\
Because $\mathbf{z}^{t+1}_{\Theta^{t+1}}$ is the $k$ best approximation of $\tilde{\mathbf{z}}^{t+1}_{\Theta_{t+1}}$ by hard thresholding besides $|\Theta_{t+1}|\leq2k$ . As a result:
\begin{eqnarray}
||\mathbf{x}_{\Theta_{t+1}}-\tilde{\mathbf{z}}^{t+1}_{\Theta_{t+1}}||\geq||\mathbf{z}^{t+1}_{\Theta_{t+1}}-\tilde{\mathbf{z}}^{t+1}_{\Theta_{t+1}}||
\end{eqnarray}

Therefore, (6.2) can be transformed as:
\begin{eqnarray}
||\mathbf{h}^{t+1}||\leq2||\mathbf{x}_{\Theta_{t+1}}-\tilde{\mathbf{z}}^{t+1}_{\Theta_{t+1}}||,
\end{eqnarray}
further, we plug the expression of $\tilde{\mathbf{z}}_{\Theta_{t+1}}^{t+1}$ into (6.4). As a result, we get the inequality below:
\begin{eqnarray}
||\mathbf{h}^{t+1}||\leq2||\mathbf{z}^{t}_{\Theta_{t+1}}-\frac{\mu_t}{\phi^2}\nabla f(\mathbf{z}^{t})_{\Theta_{t+1}}-\mathbf{x}_{\Theta_{t+1}}||=2||\mathbf{h}^{t}_{\Theta_{t+1}}-\frac{\mu_t}{\phi^2}\nabla f(\mathbf{z}^{t})_{\Theta_{t+1}}||.
\end{eqnarray}
Because
\begin{eqnarray}
\nabla f(\mathbf{z}^t)_{\Theta_{t+1}}&=&\frac{2}{m} \sum_{i=1}^{m}\big(|\mathbf{a}^{T}_i\mathbf{z}^t|^2-|\mathbf{a}^{T}_i\mathbf{x}|^2\big)\mathbf{a}_{i,\Theta_{t+1}}\mathbf{a}_i^{T}\mathbf{z}^t\nonumber\\
&=&\frac{2}{m} \sum_{i=1}^{m}\big(|\mathbf{a}^{T}_i\mathbf{z}^t|^2-|\mathbf{a}^{T}_i\mathbf{x}|^2\big)\mathbf{a}_{i,\Theta_{t+1}}\mathbf{a}_i^{T}(\mathbf{h}^t+\mathbf{x})\big)\nonumber\\
&=&\frac{2}{m}\sum_{i=1}^{m}\big(2(\mathbf{a}_i^{T}\mathbf{x})^2\mathbf{a}_i^{T}\mathbf{h}^t+3\mathbf{a}_i^{T}\mathbf{x}(\mathbf{a}_i^{T}\mathbf{h}^t)^2+(\mathbf{a}_i^{T}\mathbf{h}^t)^3\big)\mathbf{a}_{i,\Theta_{t+1}}.
\end{eqnarray} 
Plugging (6.6) into (6.5), we will get:
\begin{eqnarray}
||\mathbf{h}^{t+1}||&\leq&2||\mathbf{h}^{t}_{\Theta_{t+1}}-\frac{2\mu_t}{m\phi^2}\sum_{i=1}^{m}\big(2(\mathbf{a}_i^{T}\mathbf{x})^2\mathbf{a}_i^{T}\mathbf{h}^t+3\mathbf{a}_i^{T}\mathbf{x}(\mathbf{a}_i^{T}\mathbf{h}^t)^2+(\mathbf{a}_i^{T}\mathbf{h}^t)^3\big)\mathbf{a}_{i,\Theta_{t+1}}||\nonumber\\
&\leq&2||\mathbf{h}^{t}_{\Theta_{t+1}}-\frac{2\mu_t}{m\phi^2}\sum_{i=1}^{m}2(\mathbf{a}_i^T\mathbf{x})^2\mathbf{a}_i^{T}\mathbf{h}^t\mathbf{a}_{i,\Theta_{t+1}}||\nonumber\\
&~&+2||\frac{6\mu_t}{m\phi^2}\sum_{i=1}^{m}\mathbf{a}_i^{T}\mathbf{x}(\mathbf{a}_i^{T}\mathbf{h}^t)^2\mathbf{a}_{i,\Theta_{t+1}}||+2||\frac{2\mu_t}{m\phi^2}\sum_{i=1}^{m}(\mathbf{a}_i^{T}\mathbf{h}^t)^3\mathbf{a}_{i,\Theta_{t+1}}||
\end{eqnarray}
We can split $\mathbf{a}_i^{T}\mathbf{h}^t\mathbf{a}_{i,\Theta_{t+1}}$ into two parts:
\begin{eqnarray}
\mathbf{a}_i^{T}\mathbf{h}^t\mathbf{a}_{i,\Theta_{t+1}}=\mathbf{a}_{i,\Theta_{t+1}}^{T}\mathbf{h}^t_{\Theta_{t+1}}\mathbf{a}_{i,\Theta_{t+1}}+\mathbf{a}_{i,\Theta_{t}\setminus\Theta_{t+1}}^{T}\mathbf{h}^t_{\Theta_{t}\setminus\Theta_{t+1}}\mathbf{a}_{i,\Theta_{t+1}}
\end{eqnarray}
As a result:
\begin{eqnarray}
||\mathbf{h}^{t+1}||
&\leq&2||\mathbf{h}^{t}_{\Theta_{t+1}}-\frac{2\mu_t}{m\phi^2}\sum_{i=1}^{m}2(\mathbf{a}_i^{T}\mathbf{x})^2\mathbf{a}_{i,\Theta_{t+1}}^{T}\mathbf{h}^t_{\Theta_{t+1}}\mathbf{a}_{i,\Theta_{t+1}}||\nonumber\\
&~&+2||\frac{2\mu_t}{m\phi^2}\sum_{i=1}^{m}2(\mathbf{a}_i^{T}\mathbf{x})^2\mathbf{a}_{i,\Theta_{t}\setminus\Theta_{t+1}}^{T}\mathbf{h}^t_{\Theta_{t}\setminus\Theta_{t+1}}\mathbf{a}_{i,\Theta_{t+1}}||\nonumber\\
&~&+2||\frac{6\mu_t}{m\phi^2}\sum_{i=1}^{m}\mathbf{a}_i^{T}\mathbf{x}(\mathbf{a}_i^{T}\mathbf{h}^t)^2\mathbf{a}_{i,\Theta_{t+1}}||+2||\frac{2\mu_t}{m\phi^2}\sum_{i=1}^{m}(\mathbf{a}_i^{T}\mathbf{h}^t)^3\mathbf{a}_{i,\Theta_{t+1}}||\nonumber\\
&~&:=2P_1+\frac{2\mu_t}{\phi^2} P_2+\frac{12\mu_t}{\phi^2}P_3+\frac{4\mu_t}{\phi^2}P_4.
\end{eqnarray}
It's suffice to bound for $P_1$, $P_2$, $P_3$, $P_4$.\\

\noindent\textbf{Bound for $P_1$}\\

\begin{eqnarray}
&~&||\mathbf{h}^{t}_{\Theta_{t+1}}-\frac{2\mu_t}{m\phi^2}\sum_{i=1}^{m}2(\mathbf{a}_i^{T}\mathbf{x})^2\mathbf{a}_{i,\Theta_{t+1}}^{T}\mathbf{h}^t_{\Theta_{t+1}}\mathbf{a}_{i,\Theta_{t+1}}||\nonumber\\
&=&||(\mathbf{I}-\frac{2\mu_t}{m\phi}\sum_{i=1}^{m}2(\mathbf{a}_i^{T}\mathbf{x})^2\mathbf{a}_{i,\Theta_{t+1}}\mathbf{a}_{i,\Theta_{t+1}}^T)\mathbf{h}^{t}_{\Theta_{t+1}}||\nonumber\\
&\leq&||(\mathbf{I}-\frac{2\mu_t}{m\phi^2}\sum_{i=1}^{m}2(\mathbf{a}_i^T\mathbf{x})\mathbf{a}_{i,\Theta_{t+1}}\mathbf{a}_{i,\Theta_{t+1}}^T)||||\mathbf{h}^{t}_{\Theta_{t+1}}||\nonumber\\
&\leq&max\{1-4\frac{\mu_t}{\phi^2}\underline{\lambda},4\frac{\mu_t}{\phi^2}\overline{\lambda}-1\}||\mathbf{h}^{t}_{\Theta_{t+1}}||
\end{eqnarray}
Where $\overline{\lambda}$ and $\underline{\lambda}$ is the largest eigenvalue and smallest eigenvalue of the matrix $\frac{1}{m}\sum_{i=1}^{m}(\mathbf{a}_i^{T}\mathbf{x})^2\mathbf{a}_{i,\Theta_{t+1}}\mathbf{a}_{i,\Theta_{t+1}}^T$. The two inequalities above can be deduced by the definition of the spectral norm of the matrix.\\
Then we will bound the $\underline{\lambda}$ and $\overline{\lambda}$ respectively. Because $S^*\subset\Theta_{t+1}$, thus,
\begin{eqnarray}
\frac{1}{m}\sum_{i=1}^{m}(\mathbf{a}_i^T\mathbf{x})^2\mathbf{a}_{i,\Theta_{t+1}}\mathbf{a}_{i,\Theta_{t+1}}^T=\frac{1}{m}\sum_{i=1}^{m}(\mathbf{a}_{i,\Theta_{t+1}}^{T}\mathbf{x}_{\Theta_{t+1}})^2\mathbf{a}_{i,\Theta_{t+1}}\mathbf{a}_{i,\Theta_{t+1}}^T
\end{eqnarray}
From corollarry 5.35\cite{Vershynin2011Introduction} and Lemma 6.3\cite{sun2016geometric}, we have:
\begin{eqnarray}
\overline{\lambda}=\lambda_{max}(\frac{1}{m}\sum_{i=1}^{m}(\mathbf{a}_{i,\Theta_{t+1}}^{T}\mathbf{x}_{\Theta_{t+1}})^2\mathbf{a}_{i,\Theta_{t+1}}\mathbf{a}_{i,\Theta_{t+1}}^T)\leq (3+\delta_1)||\mathbf{x}||^2,
\end{eqnarray}
Where the inequality holds for any fixed $\delta_1\in(0,1)$, $t_1>0$ with probability $1-c_1\delta_1^{-2}m^{-1}-c_2\mathrm{exp}(-c_3\delta_1^2m/\mathrm{log}m)$ provided that $m\geq C(\delta_1)(k\mathrm{log}k)$.\\
On the another hand, using Lemma 6.3 in \cite{candes2015phase}, we can also get the inequality below:
\begin{eqnarray}
\frac{1}{m}\sum_{i=1}^{m}((\mathbf{a}^{T}_{i,\Theta_{t+1}}\mathbf{x}_{\Theta_{t+1}})^2\mathbf{a}^{T}_{i,\Theta_{t+1}}\mathbf{h}_{\Theta_{t+1}}^t)^2\geq(3-\delta_1)||\mathbf{x}||^2||\mathbf{h}_{\Theta_{t+1}}||^2
\end{eqnarray}\\
Thus we have:
\begin{eqnarray}
\underline{\lambda}=\lambda_{min}(\frac{1}{m}\sum_{i=1}^{m}(\mathbf{a}_{i,\Theta_{t+1}}^{T}\mathbf{x}_{\Theta_{t+1}})^2\mathbf{a}_{i,\Theta_{t+1}}\mathbf{a}_{i,\Theta_{t+1}}^T)\geq(3-\delta_1)||\mathbf{x}||^2
\end{eqnarray}
Above all, $P_1\leq max\big\{1-4\frac{\mu_t}{\phi^2}(3-\delta_1)||\mathbf{x}||^2,4\frac{\mu_t}{\phi^2}(3+\delta_1)||\mathbf{x}||^2-1\big\}\big|\big|\mathbf{h}^{t}_{\Theta_{t+1}}\big|\big|$\\

\noindent\textbf{Bound for $P_2$}

\begin{eqnarray}
&~&||\frac{1}{m}\sum_{i=1}^{m}(\mathbf{a}_i^{T}\mathbf{x})^2\mathbf{a}_{i,\Theta_{t}\setminus\Theta_{t+1}}^{T}\mathbf{h}^t_{\Theta_{t}\setminus\Theta_{t+1}}\mathbf{a}_{i,\Theta_{t+1}}||\nonumber\\
&\leq&||\mathbf{x}\mathbf{x}^{T}+||\mathbf{x}||^2\mathbf{I}-\frac{1}{m}\sum_{i=1}^{m}(\mathbf{a}_{i,\Theta_{t+1}}^{T}\mathbf{x}_{\Theta_{t+1}})^2\mathbf{a}_{i,\Theta_{t}\cup\Theta_{t+1}}\mathbf{a}_{i,{\Theta_{t}\cup\Theta_{t+1}}}^T||||\mathbf{h}^t_{\Theta_{t}\setminus\Theta_{t+1}}||\nonumber\\
&\leq&\delta_2||\mathbf{x}||^2||\mathbf{h}^t_{\Theta_{t}\setminus\Theta_{t+1}}||.
\end{eqnarray}
This holds with probability $1-2\mathrm{exp}(-C_2(\delta_2)m)$ provided that $m\geq C_3(\delta_2)3k\mathrm{log}(3k)$. Namely, the ratio $m/3k$ betweent the number of the measurements and the sparsity of $\mathbf{x}$ will exceed a sufficiently large constant.\\
\indent
The second inequality is derived from Lemma 6.3\cite{sun2016geometric} and Proposition 3.2 \cite{Needell2008CoSaMP} which is called approximate orthogonality.\\
\indent
As a result, we can conclude that $P_2\leq\frac{2\mu_t\delta_2}{\phi^2}||\mathbf{x}||^2||\mathbf{h}^{t}_{\Theta_t\setminus\Theta_{t+1}}||$\\
\indent
Now, we will consider the last two terms. The skills to bound this two terms is the same. Thus we only show the details of bounding $P_3$.\\

\noindent\textbf{Bound for $\mathbf{P_3}$}\\

\indent
Define $\mathbf{A}_{\Theta_{t+1}}=[\mathbf{a}_{1,\Theta_{t+1}},...,\mathbf{a}_{m,\Theta_{t+1}}]$, and $\mathbf{v}=[v_1,...,v_m]^{T}$ where $v_i=\mathbf{a}_i^{T}\mathbf{x}(\mathbf{a}_i^{T}\mathbf{h}^t)^2$. As a result:
\begin{eqnarray}
||\frac{1}{m}\sum_{i=1}^{m}\mathbf{a}_i^{T}\mathbf{x}(\mathbf{a}_i^{T}\mathbf{h}^t)^2\mathbf{a}_{i,\Theta_{t+1}}||^2&=&||\frac{1}{m}\mathbf{A}_{\Theta_{t+1}}\mathbf{v}||\nonumber\\
&\leq&||\frac{1}{\sqrt{m}}\mathbf{A}_{\Theta_{t+1}}||||\frac{1}{\sqrt{m}}\mathbf{v}||,
\end{eqnarray}
By the standard matrix concentration results, for any fixed $\varepsilon_1>0$, the largest singluar value of $\mathbf{A}_{\Theta_{t+1}}$ satisfied $s_{max}(\mathbf{A}_{\Theta_{t+1}})\leq(1+\varepsilon_1)\sqrt{m}$ with probability at least $1-2\mathrm{exp}(-C_3(\varepsilon_1)m)$ provided $m\geq C_0k$ for some large constant $C_0$. Thus,
\begin{eqnarray}
||\frac{1}{m}\sum_{i=1}^{m}\mathbf{a}_i^{T}\mathbf{x}(\mathbf{a}_i^{T}\mathbf{h}^t)^2\mathbf{a}_{i,\Theta_{t+1}}||^2\leq(1+\varepsilon_1)||\frac{1}{\sqrt{m}}\mathbf{v}||
\end{eqnarray}
can be held with high probability.\\
Next, we will bound $||\frac{1}{\sqrt{m}}\mathbf{v}||$. Note that:
\begin{eqnarray}
||\frac{1}{\sqrt{m}}\mathbf{v}||^2=\frac{1}{m}\sum_{i=1}^{m}(\mathbf{a}_i^{T}\mathbf{x})^2(\mathbf{a}_i^{T}\mathbf{h}^t)^4.
\end{eqnarray}
For $\frac{1}{m}\sum_{i=1}^{m}(\mathbf{a}_i^{T}\mathbf{x})^2(\mathbf{a}_i^T\mathbf{h}^t)^4$, with Holder inequality we will have:
\begin{eqnarray}
\frac{1}{m}\sum_{i=1}^{m}(\mathbf{a}_i^{T}\mathbf{x})^2(\mathbf{a}_i^{T}\mathbf{h}^t)^4&\leq&\frac{1}{m}(\sum_{i=1}^{m}(\mathbf{a}_i^{T}\mathbf{x})^6)^{\frac{1}{3}}(\sum_{i=1}^{m}(\mathbf{a}_i^{T}\mathbf{h}^t)^6)^{\frac{2}{3}}\nonumber\\
&\leq&\frac{1}{m}((15m)^{1/6}+k^{\frac{1}{2}}+(2\mathrm{log}m)^{1/3}||\mathbf{x}||^2||\mathbf{h}^t||^4\nonumber\\
&\leq&17\delta_3^2||\mathbf{x}||^4||\mathbf{h}^t||^2.
\end{eqnarray}
The second inequality above is derived from Lemma A.5\cite{Cai2016Optimal} which holds with $m\geq C_4k$ with probability at least $1-2/m$. The last inequality is derived from lemma 2.\\
\indent
So
\begin{eqnarray}
||\frac{1}{m}\sum_{i=1}^{m}\mathbf{a}_i^{T}\mathbf{x}(\mathbf{a}_i^{T}\mathbf{h}^t)^2\mathbf{a}_{i,\Theta_{t+1}}||^2\leq\sqrt{17}\delta_3||\mathbf{x}||^2||\mathbf{h}^t||.
\end{eqnarray} 
Which holds with probability $1-2\mathrm{exp}(C_5m)$ provided $m\geq C_6k$.\\
With the same ideas, we give the bound for $P_4$:
\begin{eqnarray}
\mathbf{P}_4=||\frac{1}{m}\sum_{i=1}^{m}(\mathbf{a}_i^{T}\mathbf{h}^t)^3\mathbf{a}_{i,\Theta_{t+1}}||\leq\sqrt{17}\delta_3^2||\mathbf{x}||^2||\mathbf{h}^t||
\end{eqnarray}\\
Combine those bounds together, we will have:
\begin{eqnarray}
||\mathbf{h}^{t+1}||&\leq&2P_1+\frac{2\mu_t}{\phi^2} P_2+\frac{12\mu_t}{\phi^2}P_3+\frac{4\mu_t}{\phi^2}P_4\nonumber\\
&\leq&2max\big\{1-2\frac{\mu_t}{\phi^2}(3-\delta_1)||\mathbf{x}||^2,2\frac{\mu_t}{\phi^2}(3+\delta_1)||\mathbf{x}||^2-1\big\}\big|\big|\mathbf{h}^{t}_{\Theta_{t+1}}\big|\big|\nonumber\\
&~&+\frac{2\mu_t\delta_2}{\phi^2}||\mathbf{x}||^2||\mathbf{h}^t_{\Theta_t\setminus\Theta_{t+1}}||+\frac{12\sqrt{17}\delta_3\mu_t}{\phi^2}||\mathbf{x}||^2||\mathbf{h}^t||+\frac{4\sqrt{17}\delta_3^2\mu_t}{\phi^2}||\mathbf{x}||^2||\mathbf{h}^t||.
\end{eqnarray}
From Lemma 6.2\cite{Cai2016Optimal}, we will have:
\begin{eqnarray}
1-\delta_4\leq\frac{||\mathbf{x}||^2}{\phi^2}\leq1+\delta_4
\end{eqnarray}
with probability at least $1-\frac{3}{m}$ as long as $\frac{m}{\mathrm{log}m}$ exceeding a sufficiently large constant $C_7$.\\
Because $\Theta_t\cap(\Theta_{t}\setminus\Theta_{t+1})=\varnothing$, $||\mathbf{h}^t_{\Theta_{t+1}}||^2+||\mathbf{h}^t_{\Theta_{t}\setminus\Theta_{t+1}}||^2=||\mathbf{h}^t_{\Theta_t}||^2\leq||\mathbf{h}^t||^2$. Utilizing $(a+b)^2\leq2(a^2+b^2)$, as a result we have
\begin{eqnarray}
||\mathbf{h}^t_{\Theta_{t+1}}||+||\mathbf{h}^t_{\Theta_{t}\setminus\Theta_{t+1}}||\leq\sqrt{2}||\mathbf{h}^t||.
\end{eqnarray} 
So,
\begin{eqnarray}
||\mathbf{h}^{t+1}||&\leq&\Big((\sqrt{2}max\Big\{2max\big\{1-4{\mu_t}(1-\delta_4)(3-\delta_1),4\mu_t(1+\delta_4)(3+\delta_1)-1\big\}\nonumber\\
&~&,2\mu_t\delta_2(1+\delta_4)\Big\}+(12\sqrt{17}+4\sqrt{17}\delta_3)(1+\delta_4)\mu_t\delta_3\Big)||\mathbf{h}^t||
\end{eqnarray}
Specifically, let $\delta_1$, $\delta_2$, $\delta_3$ and $\delta_4$ be sufficiently small, we can obtain the feasible range for $\mu_t$: 
\begin{eqnarray}
\nu_1\leq\mu_t\leq\nu_2,
\end{eqnarray}
where $\nu_1$, $\nu_2$ are constants.\\
So,
\begin{eqnarray}
||\mathbf{h}^{t+1}||\leq(1-\nu)||\mathbf{h}^t||.
\end{eqnarray}
where $\nu$ is a constant. $0<\nu<1$.\\
Specifically, when $\delta_1=\delta_2=\delta_4=0.001$, $\delta_3=0.05$, let $\mu_t=0.1$, we have:\\
\begin{eqnarray}
||\mathbf{h}^{t+1}||\leq 0.81||\mathbf{h}^{t}||.
\end{eqnarray}
Then we conclude our proof.
\end{document}